\documentclass{amsart}
\usepackage{amssymb}

\newtheorem{theorem}{Theorem}

\numberwithin{equation}{section}

\newcommand{\lbeq}[1]  {\label{eq: #1}}
\newcommand{\refeq}[1] {(\ref{eq: #1})}

\newcommand{\eqarray}   {\begin{eqnarray}} 
\newcommand{\enarray}   {\end{eqnarray}}

\newcommand{\RBbb} {{\mathbb R}}
\newcommand{\ZBbb} {{ {\mathbb Z} }}

\begin{document}

\title  {
        The incipient infinite cluster in 
	high-dimensional percolation
        }

\author{Takashi Hara}
\address{Department of Applied Physics,
        Tokyo Institute of Technology,
        Oh-Okayama, Meguro-ku, Tokyo 152,
        Japan}
\email{hara@ap.titech.ac.jp} 

\author{Gordon Slade}
\address{Department of Mathematics and Statistics, 
        McMaster University,
        Hamilton, ON,  Canada L8S 4K1}
\email{slade@mcmaster.ca}


\date{May 22, 1998}
\thanks{To appear in Electronic Research Announcements of the AMS, 
Volume 4, (1998).  http://www.ams.org/era/}


\begin{abstract}
We announce our recent proof that, for independent bond percolation in 
high dimensions, the scaling limits of the incipient infinite cluster's 
two-point and three-point functions are those of integrated 
super-Brownian excursion (ISE).  The proof uses an extension of the lace 
expansion for percolation.
\end{abstract}

\maketitle

\section{Introduction}
\label{sec-intro}

Percolation has received much attention in mathematics and in physics,
as a simple model of a phase transition.
To describe the phase transition, associate to each {\em bond}\/
$\{x,y\}$ ($x,y \in {\ZBbb}^d$, separated by unit Euclidean distance)  
a Bernoulli random variable $n_{\{x,y\}}$ taking the value 1 with probability
$p$ and the value
0 with probability $1-p$.  These random variables are independent,
and $p$ is a control parameter in $[0,1]$.  A bond $\{x,y\}$ 
is said to be {\em occupied}\/ if $n_{\{x,y\}}=1$, and {\em vacant}\/ if
$n_{\{x,y\}}=0$.  The control parameter $p$ is thus the density of occupied
bonds in the infinite lattice ${\ZBbb}^d$.  The percolation phase
transition is the fact that, for $d \geq 2$, 
there is a critical value $p_c=p_c(d) \in (0,1)$,
such that for $p<p_c$ there is with probability 1 no infinite connected
cluster of occupied bonds, whereas for $p>p_c$ there is with probability
1 a unique infinite connected cluster of occupied bonds (percolation occurs).

There is now a considerable mathematical understanding of the
subcritical regime $p<p_c$ and the supercritical regime $p>p_c$ 
\cite{Grim89,Grim96,Hugh96,Kest82},
but the rich universal behaviour at the critical point $p=p_c$ remains
largely the province of physics rather than mathematics.  In particular,
there is still no general proof of the widely-accepted statement that
there is no infinite cluster when $p=p_c$.  This has been proved only
for $d=2$ \cite{Grim89} (and references therein),
$d \geq 19$ \cite{BA91,HS90a,HS94}, 
and for $d>6$ for sufficiently spread-out models \cite{BA91,HS90a}
of the type described below.  We focus in this paper on the 
high-dimensional case, where the absence of percolation at $p_c$ has been
established.

The percolation phase transition presents a picture where at $p_c$ there
are extensive connections present, on all length scales, but no infinite
cluster.  However, the
slightest increase in $p$ will lead to the formation of an infinite
cluster.  This inchoate state of affairs at $p_c$ is often represented
by an appeal to the notion of the ``incipient infinite cluster.''

The incipient infinite cluster has been defined in 2-dimensional models
as an infinite cluster in ${\ZBbb}^2$ constructed by an appropriate
limiting process \cite{Kest86}, or by introducing an inhomogeneity 
\cite{CCD87}.  
We will approach the incipient infinite cluster
from a different perspective.  Rather than attempting to construct
an infinite object on the lattice, we instead take a scaling limit
of increasingly large but finite clusters, at $p=p_c$.  This involves shrinking
the lattice spacing as a function of the cluster size $n$ in such
a way as to produce a nontrivial random subset of the continuum ${\RBbb}^d$, 
in the limit $n \to \infty$.
This is analogous to the procedure by which Brownian motion on the time 
interval $[0,1]$ can be constructed as a limit of an increasingly long 
lattice random walk.
The appropriate spatial scaling of the lattice is presumably $n^{-1/D_H}$,
where $D_H$ corresponds to the Hausdorff
dimension of the incipient infinite cluster.  For $d>6$, it is believed
that $D_H=4$, and accordingly we 
will scale down the lattice spacing by $n^{1/4}$.

Let $x \in {\RBbb}^d$ be fixed. 
We prove that in sufficiently high dimensions, the probability that
a site $\lfloor xn^{1/4} \rfloor \in {\ZBbb}^d$ 
is connected to the origin in a cluster of
size $n$, corresponds, in the scaling limit $n \to \infty$, to the 
mean mass density function of integrated super-Brownian
excursion (ISE) at $x$.  
This will be stated more precisely in Theorem~\ref{thm-2pt} below.
ISE is a measure-valued stochastic process, representing a
continuous-time branching process in which
branching occurs on all (arbitrarily short) length 
scales \cite{Aldo93,DP97}, and forms
a basic example in the theory of superprocesses.  Its state at time $t \geq 0$
is a random measure on ${\RBbb}^d$ which, integrated over $t$, is a random
probability measure on ${\RBbb}^d$.  For $d >4$, the support of this random
probability measure almost surely has Hausdorff dimension $4$ \cite{DP97}.
The support of the ISE probability measure is almost surely 
a compact random subset of ${\RBbb}^d$, but this corresponds
to an infinite lattice object.

We also prove that in sufficiently high dimensions, 
the probability that the origin is connected to 
sites $\lfloor xn^{1/4}\rfloor$ and $\lfloor yn^{1/4} \rfloor$ 
($x,y \in {\RBbb}^d$) corresponds, 
in the scaling limit, to the ISE mean joint mass density at
$(x,y)$.  A precise statement will be given
in Theorem~\ref{thm-3pt} below.

We conjecture that the scaling limit of the incipient infinite cluster
is ISE for all dimensions $d>6$.  The {\em upper critical dimension}\/ 6
has long been identified as the dimension above which the behaviour of
percolation models near $p_c$ no longer exhibits the dimension-dependence
typical of lower dimensions, and adopts behaviour associated with percolation
on trees.  Our results mentioned above for the two- and three-point functions
are restricted to sufficiently high dimensions
(we have not computed {\em how}\/ high is sufficient), rather than to
$d>6$, in part because
we use an expansion method, the lace expansion, for which the inverse
dimension is the small parameter ensuring convergence.

There is an alternate small parameter that has been used in lace expansion
methods in the past, which removes the need for the spatial dimension to
serve as a small parameter, and allows for a control of {\em all}\/
dimensions above the  
upper critical dimension.  This involves the introduction of spread-out models,
in which the nearest-neighbour bonds used above are enriched to a set of
bonds of the form $\{x,y\}$ with $0 <\|x-y\| \leq L$, where the norm is,
for example, the supremum norm.  
Again we define Bernoulli random variables $n_{\{x,y\}}$ for each bond,
just as was done for the nearest-neighbour model.
We take $L$ large, with $L^{-1}$ serving as 
a small parameter to make the lace expansion converge.
The conventional wisdom (still unproved in general), 
and an assertion of the hypothesis of universality, is that in any
dimension $d$ the spread-out models have identical critical behaviour
for all finite $L \geq 1$, and for any choice of norm which respects the 
lattice symmetries.  This is analogous to the fact that diverse
lattice random walks have the same Brownian scaling limit.

At present, our method is not adequate to prove that the scaling limits
of the probability of a connection of two points, or three points, is
the corresponding ISE density for sufficiently spread-out models in all
dimensions $d>6$.  This is due to a difficulty, associated with the
fact that ISE has self-intersections in dimension less than 8, which currently 
prevents us from handling dimensions 7 and 8 in such detail.  However,
as we will indicate below in Theorem~\ref{thm-taubd}, 
we are able to give some indication that
ISE may be relevant to the scaling limit of the incipient infinite cluster,
for sufficiently spread-out models in all dimensions $d>6$.

The study of the scaling limit of the incipient infinite cluster is
basic in the analysis of the continuum limit of critical percolation.
Above six dimensions, work in this direction 
has been carried out by Aizenman \cite{Aize97} (see also 
\cite{BCKS98a,BCKS98b}).
Aizenman's results, which are based on
the assumption (as yet, unproved) that at $p_c$ the probability of
a distant site $x$ being connected to the origin is 
comparable to $|x|^{2-d}$, are complementary to ours.  
In particular, a picture is described in \cite{Aize97} for percolation
on a lattice with $d>6$ and with small spacing $a$,
where in a window of fixed size in the continuum, 
the largest clusters have size of order $a^{-4}$ and there are of order
$a^{6-d}$ clusters of this size.  
Our results suggest that
for $d>6$ a cluster of size $n=a^{-4}$ in a lattice with spacing
$a=n^{-1/4}$ will typically be an ISE cluster, in the scaling limit.

\section{The results}
\label{sec-results}
\setcounter{equation}{0}

Consider independent 
bond percolation (nearest-neighbour or spread-out) with $p=p_c$.
Let $C(0)$ denote the random set of sites connected to $0$ by a path
consisting of occupied bonds, and let $|C(0)|$ denote its cardinality.  Let
\begin{equation}
        \tau(x; n) = P(C(0) \ni x , |C(0)|=n )
\end{equation}
denote the probability at $p_c$ 
that the origin is connected to $x$ by a cluster containing $n$ sites.
Then
\begin{equation}
        q_n(x) = \frac{\tau(x; n)}{\sum_{x \in {\ZBbb}^d} \tau(x; n)}
        = \frac{\tau(x; n)}{n P( |C(0)|=n ) }
\end{equation}
defines a probability measure on ${\ZBbb}^d$ proportional to the conditional
probability at $p_c$ that a cluster of size $n$ contains $x$.

For $k \in {\RBbb}^d$, define
\begin{equation}
\lbeq{Ak}
        \hat{A}^{(2)}(k) = \int_0^\infty t e^{-t^2/2} e^{-k^2 t/2} dt .
\end{equation}
This is the Fourier integral transform of the mean mass density function
\begin{equation}
	A^{(2)}(x) = \int_0^\infty t e^{-t^2/2} (2\pi t)^{-d/2} e^{-x^2/2t} dt 
\end{equation}
of ISE;
for a discussion of this formula, see 
\cite{Aldo93,DS97,Aldo93a,LeGa93}.
The following theorem shows that in the scaling limit, the two-point function
of the incipient infinite cluster is the two-point function of ISE, in
sufficiently high dimensions.  In its statement, we use the discrete
Fourier transform $\hat{f}(k) = \sum_{x \in {\ZBbb}^d} f(x) e^{ik\cdot x}$
($k \in [-\pi,\pi]^d$), for a summable function $f$ on ${\ZBbb}^d$.

\begin{theorem}
\label{thm-2pt}
Fix $k \in \RBbb^d$ and any $\epsilon \in (0, \frac{1}{2})$.
For the nearest-neighbour model with $d$ sufficiently large and $p=p_c$, 
there are $d$-dependent positive constants $C,D$ such that
\begin{equation}
        \hat{\tau}(kD^{-1}n^{-1/4};n) = 
        \frac{C}{\sqrt{8\pi n}} \hat{A}^{(2)}(k)[1+O(n^{-\epsilon})].
\end{equation}
In particular, 
\begin{equation}
\lbeq{delta2}
        P(|C(0)|=n) 
        = \frac{1}{n} \hat{\tau}(0;n)
        = \frac{C}{\sqrt{8\pi}n^{3/2}}
        [1+O(n^{-\epsilon})],
\end{equation}
and
\begin{equation}
\lbeq{qn2}
        \lim_{n \to \infty} \hat{q}_n(kD^{-1}n^{-1/4}) = \hat{A}^{(2)}(k).
\end{equation}
\end{theorem}

Equation~\refeq{delta2} asserts that $\delta =2$, where $\delta$ is
the critical exponent in the conjectured relation $P(|C(0)|=n) \approx
n^{-1-1/\delta}$.
Equation~\refeq{qn2} can be interpreted as asserting that
in the scaling limit the distribution of a site 
$\lfloor x D n^{1/4} \rfloor$ in the cluster
of the origin, conditional on the cluster being of size $n$,
is the distribution of a point in ISE.

We now consider the three-point function.  Let $\tau^{(3)}(x,y; n)$ denote
the probability, at $p_c$, 
that the origin is connected to $x$ and $y$ and that the
cluster of the origin contains exactly $n$ sites.  
For $k, l \in [-\pi,\pi]^d$,
define
\begin{equation}
\lbeq{t3def}
        \hat{\tau}^{(3)}(k,l;n)
        = \sum_{x,y \in {\ZBbb}^d} \tau^{(3)}(x,y; n) e^{ik\cdot x}
         e^{il\cdot y}.
\end{equation}
We define a probability measure on ${\ZBbb}^{2d}$ by
\begin{equation}
        q_n^{(3)}(x,y) = \frac{\tau^{(3)}(x,y; n)}
        {\sum_{x,y \in {\ZBbb}^d}\tau^{(3)}(x,y; n)}
        = \frac{\tau^{(3)}(x,y; n)}
        {n^2 P ( |C(0)|=n ) }.
\end{equation}
For $k,l \in {\RBbb}^d$, let $\hat{A}^{(3)}(k,l)$
denote the Fourier transform of the ISE three-point function:
\begin{equation}
\lbeq{A3k}
        \hat{A}^{(3)}(k,l) = \int_0^\infty \int_0^\infty \int_0^\infty 
        \left( \sum_{j=1}^3 t_j \right)
         e^{-(\sum_{j=1}^3 t_j)^2/2} 
        e^{- [(k+l)^2t_1 +k^2 t_2 + l^2t_3]/2} dt_1 \, dt_2 \, dt_3.
\end{equation}
Equation~\refeq{A3k} differs from the formulas of 
\cite{Aldo93,DS97} in that here we have not fixed the location
of the internal branch point.
The next theorem shows that in the scaling limit, the three-point function
of the incipient infinite cluster corresponds to that of ISE, in high
dimensions. 
The constants $C,D$ in the theorem are the same as those appearing in
Theorem~\ref{thm-2pt}.

\begin{theorem}
\label{thm-3pt}
Fix $k \in \RBbb^d$ and any $\epsilon \in (0,\frac{1}{2})$.
For the nearest-neighbour model with $d$ sufficiently large and $p=p_c$,
\begin{equation}
        \hat{\tau}^{(3)}(kD^{-1}n^{-1/4}, lD^{-1}n^{-1/4};n)
        = 
        \frac{C}{\sqrt{8\pi}} n^{1/2}
        \hat{A}^{(3)}(k,l)[1 + O(n^{-\epsilon})].
\end{equation}
In particular, 
\begin{equation}
        \lim_{n \to \infty} 
        \hat{q}_n^{(3)}(kD^{-1}n^{-1/4},lD^{-1}n^{-1/4})
        = \hat{A}^{(3)}(k,l).
\end{equation}
\end{theorem}

Note that for $(k,l)=(0,0)$, Theorem~\ref{thm-3pt} follows immediately
from Theorem~\ref{thm-2pt}, since 
$\hat{\tau}^{(3)}(0,0;n) = n^2 P(|C(0)|=n)$ and $\hat{A}^{(3)}(0,0)=1$.

As mentioned in Section~\ref{sec-intro}, for sufficiently spread-out models
in dimensions $d>6$, we have a weaker result.
In preparation for this, we define
a generating function $\Lambda_z(k)$ and coefficients $\lambda_n(k)$ by
\begin{equation}
\lbeq{Cdef}
        \Lambda_z(k) = \frac{1}{k^2 + 2^{3/2} \sqrt{1-z}} 
        = \sum_{n=0}^\infty \lambda_n(k) z^n.
\end{equation}
The square root has branch cut $[1,\infty)$, and the branch with 
$\sqrt{1-z}$ positive for $z \in (-\infty,1)$ is chosen.  
The power series has radius of convergence 1.  
By Cauchy's theorem,
\begin{equation}
\lbeq{Cint}
        \lambda_n(k) = \frac{1}{2\pi i} \oint_\Gamma \Lambda_{z}(k)
        \frac{dz}{z^{n+1}},
\end{equation}
where $\Gamma$ is a circle centred at the origin, of any radius less than 1.
An elementary computation extending \cite[Lemma~1]{DS97a} 
shows that 
\begin{equation}
\lbeq{cnasy}
        \lambda_n(kn^{-1/4})
	= \frac{1}{\sqrt{ 8\pi n}} \hat{A}^{(2)}(k) +O(n^{-3/2}),
\end{equation}
demonstrating a link between ISE and the generating function $\Lambda_z(k)$.
As we will explain in Section~\ref{sec-method}, 
this link is a key element in the proof of Theorems~\ref{thm-2pt}
and \ref{thm-3pt}.

We define the generating function
\begin{equation}
\lbeq{Mzdef}
        \tau_{z}(x) = \sum_{n=1}^\infty \tau(x;n) z^{n},
        \quad |z| \leq 1.
\end{equation}
The parameter $z$ is a complex variable.  It is not hard to show that
the Fourier transform  
$\hat{\tau}_z(k) = \sum_x \tau_{z}(x) e^{ik \cdot x}$
exists for $|z| <1$.
When $z \in [0,1]$, it is traditional to write $z = e^{-h}$, with $h$ playing
the role of a magnetic field, but since here $z$ is in general complex,
we will not adopt this notation.  
The following theorem shows that the generating function $\Lambda_z(k)$ is
relevant for sufficiently spread-out percolation models in all $d>6$,
and provides a statement, in that context, linking the incipient
infinite cluster to ISE.

\begin{theorem}
\label{thm-taubd}
For any $d>6$, there are positive constants $C,D$ (depending on $d,L$) and
an $L_0(d)$ (large), such that for $L \geq L_0(d)$,
$k \in [-\pi,\pi]^d$ and $z \in [0,1)$, 
\begin{equation}
        \hat{\tau}_{z}(k)= C \Lambda_z(Dk)[1+\epsilon(z,k)],
\end{equation}
where $|\epsilon(z,k)| \leq \epsilon_1(z) + \epsilon_2(k)$
with $\lim_{z \to 1} \epsilon_1(z) = \lim_{k \to 0} \epsilon_2(k) =0$.
\end{theorem} 

The control of the limit $z \to 1$ provides 
a somewhat different statement from 
the infra-red bound of \cite{HS90a} 
that $\eta =0$, while taking $k=0$ recovers the statement $\delta =2$
from \cite{BA91,HS90a}.  The critical exponents $\eta$ and $\delta$
appear in the relations
$\hat{\tau}_{1}(k) \approx k^{-2+\eta}$ as $k \to 0$, and
$\hat{\tau}_{z}(0) \approx (1-z)^{1/\delta-1}$ as $z \uparrow 1$, 
which are conjectured to hold in general dimensions, with $d$-dependent
values for the exponents when $d<6$.

The proof of Theorem~\ref{thm-taubd} is given in \cite{HS98b},
and Theorems~\ref{thm-2pt} and \ref{thm-3pt} are proved in \cite{HS98c}.

\section{The method}
\label{sec-method}
\setcounter{equation}{0}

The method of proof of Theorem~\ref{thm-2pt} involves showing
that in high dimensions it is possible to write, for $|z|<1$,
\begin{equation}
\lbeq{tauE}
        \hat{\tau}_{z}(k) 
        = \frac{C}{D^2k^{2} + 2^{3/2}(1-z)^{1/2}}
        + E_z(k)
	= C\Lambda_z(Dk) + E_z(k),
\end{equation}
with $E_z(k) = \sum_{n=1}^\infty e_n(k) z^n$ 
and $|e_n(kD^{-1}n^{-1/4})| \leq O(n^{-\epsilon -1/2})$.  This is
sufficient, in view of \refeq{cnasy}.  
The leading behaviour of \refeq{tauE}
corresponds to the mean-field critical exponents $\delta=2$ and $\eta=0$.
Theorem~\ref{thm-taubd} is in the same spirit as \refeq{tauE}, but does not
involve complex variables or power-law error estimates, and is
easier to establish.

For Theorem~\ref{thm-3pt}, we define $\hat{\tau}_z^{(3)}(k,l) =
\sum_{n=1}^\infty \hat{\tau}^{(3)}(k,l;n) z^n$ and show that
\begin{equation}
\lbeq{tau3E}
        \hat{\tau}^{(3)}_z(k,l) 
	=
        4 C \Lambda_z(D(k+l)) \Lambda_z(Dk) \Lambda_z(Dl)
        + E^{(3)}_z(k,l),
\end{equation}
with $E^{(3)}_z(k,l) = \sum_{n=1}^\infty e_n^{(3)}(k,l) z^n$ 
and $|e_n^{(3)}(kD^{-1}n^{-1/4},lD^{-1}n^{-1/4})| \leq 
O(n^{-\epsilon +1/2})$. 
An elementary contour integration, as in \refeq{Cint} but for a product
of $\Lambda_z$'s, then gives the theorem.
Combining \refeq{tauE} and
\refeq{tau3E}, to leading order the three-point function obeys
\begin{equation}
\lbeq{tau3fac}
	\hat{\tau}^{(3)}_z(k,l) 
	=
	4 C^{-2}
	\hat{\tau}_{z}(k+l) \hat{\tau}_{z}(k) \hat{\tau}_{z}(l) +
	\mbox{error.}
\end{equation}
This
factorization corresponds to an effective independence that is discussed
further below, and is in the spirit of a conjecture of \cite{AN84}.

Our proof of \refeq{tauE} involves an analysis of the $z$-derivative of
$\hat{\tau}_z(k)$.  This leads naturally to the study 
of $\hat{\tau}^{(3)}_z$, since
\eqarray
\lbeq{tauder}
        z \frac{d}{dz} \hat{\tau}_z(k) 
        & = &\sum_{x \in {\ZBbb}^d} e^{ik\cdot x}
          \sum_{n=1}^\infty n \tau(x;n) z^n
	\\ \nonumber 
        & = & \sum_{x,y \in {\ZBbb}^d} e^{ik\cdot x}
          \sum_{n=1}^\infty \tau^{(3)}(x,y;n) z^n 
        = \hat{\tau}^{(3)}_z (k,0).
\enarray
The study of the three-point function is thus central for our method.

Equation \refeq{tau3fac} can be understood in terms of the figure
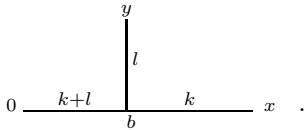
\begin{center}
\setlength{\unitlength}{0.0060000in}%
\begingroup\makeatletter\ifx\SetFigFont\undefined
\def\x#1#2#3#4#5#6#7\relax{\def\x{#1#2#3#4#5#6}}%
\expandafter\x\fmtname xxxxxx\relax \def\y{splain}%
\ifx\x\y   
\gdef\SetFigFont#1#2#3{%
  \ifnum #1<17\tiny\else \ifnum #1<20\small\else
  \ifnum #1<24\normalsize\else \ifnum #1<29\large\else
  \ifnum #1<34\Large\else \ifnum #1<41\LARGE\else
     \huge\fi\fi\fi\fi\fi\fi
  \csname #3\endcsname}%
\else
\gdef\SetFigFont#1#2#3{\begingroup
  \count@#1\relax \ifnum 25<\count@\count@25\fi
  \def\x{\endgroup\@setsize\SetFigFont{#2pt}}%
  \expandafter\x
    \csname \romannumeral\the\count@ pt\expandafter\endcsname
    \csname @\romannumeral\the\count@ pt\endcsname
  \csname #3\endcsname}%
\fi
\fi\endgroup
\begin{picture}(225,109)(105,644)
\thinlines
\put(120,660){\line( 1, 0){200}}
\put(210,660){\line( 0, 1){ 80}}
\put(105,660){\makebox(0,0)[lb]{\smash{$\scriptstyle 0$}}}
\put(330,660){\makebox(0,0)[lb]{\smash{$\scriptstyle x$}}}
\put(360,660){\makebox(0,0)[lb]{\smash{.}}}
\put(205,745){\makebox(0,0)[lb]{\smash{$\scriptstyle y$}}}
\put(210,645){\makebox(0,0)[lb]{\smash{$\scriptstyle b$}}}
\put(150,665){\makebox(0,0)[lb]{\smash{$\scriptstyle k+l$}}}
\put(215,700){\makebox(0,0)[lb]{\smash{$\scriptstyle l$}}}
\put(260,665){\makebox(0,0)[lb]{\smash{$\scriptstyle k$}}}
\end{picture}
\end{center}
The connections in the figure represent edge-disjoint
connections by occupied paths in the cluster of the origin, and
the branch point $b$ is not uniquely defined.
In high dimensions, 
the three parts of the diagram, corresponding to the connections
$0 \to b$, $b \to x$, $b \to y$, can be regarded, to leading
order, as effectively {\em independent}.  This independence is not
exact, but rather correction terms involving the triangle diagram
of \cite{AN84}, and related diagrams, 
give rise to the renormalized vertex factor $4C^{-2}$ 
appearing in \refeq{tau3fac}.

The establishment of this effective independence is at
the heart of the lace expansion method of \cite{HS90a} 
(see also \cite{HS94,MS93}), and allows for the demonstration of  
the independence of the connections $0 \to b$ and $b \to x$,
without the presence of the variable $y$.  A second expansion is
required to demonstrate the effective independence of the connection
$b \to y$.  This is technically involved, but is conceptually 
similar to the first expansion.  Such a double expansion has been
carried out previously in the analysis of lattice trees in high dimensions
\cite{HS92c}, and was used to prove results similar to Theorems~\ref{thm-2pt}
and \ref{thm-3pt} in that context \cite{DS97a,DS97}.

The previous development of the lace expansion for percolation was
restricted to the case $z=1$ \cite{HS90a}.  
Working with general $z$ provides new
difficulties to overcome.  For general $z$, we generate the expansion
using a probabilistic interpretation, valid for positive $z$, that was
used, e.g., in \cite{AB87}.  In this interpretation, the sites in
${\ZBbb}^d$ are declared to be ``not green'' with probability $z \in [0,1]$
and ``green'' with probability $1-z$.  The site variables are
independent, and independent of the bond variables.  Then, for $z \in [0,1]$,
$\tau_z(x)$ can be interpreted as the probability that the origin is
connected to $x$ but is not connected to any green site.  This green-free
condition on the two-point function necessitates major revision of the
lace expansion methodology.  The expansion can be extended 
from positive to complex $z$ via analyticity.

\section{The backbone}

ISE can be understood as a process evolving in time, and it is
of interest to interpret our results in terms of a time variable.
For this, we introduce the notion of the {\em backbone}\/ of a cluster
containing two sites $x$ and $y$.  
We define the backbone, which depends on $x,y$, 
to consist of those sites $u \in C(x)$ for
which there are edge-disjoint
paths consisting of occupied bonds from $x$ to $u$
and from $u$ to $y$.   
The backbone thus consists of connections
from $x$ to $y$, with all ``dangling ends'' removed.

We believe it would be of interest to attempt to extend our methods,
in combination with the methods of \cite{DS97a}, 
to prove that (for high dimensions) in a cluster of size
$n$, a backbone joining sites $\lfloor xn^{1/4}\rfloor$ and 
$\lfloor yn^{1/4}\rfloor$ ($x,y \in {\RBbb}^d$)
typically consists of $O(n^{1/2})$ sites and converges in the scaling limit
to a Brownian path, with the Brownian time variable corresponding to
distance along the backbone.  Such a study has not been carried out
for percolation, but
an analogous result has been proved for high-dimensional lattice trees in
\cite[Theorem~1.2]{DS97a}.  In this interpretation, the integration
variables $t$ and $t_i$ appearing in $\hat{A}^{(2)}(k)$ and $\hat{A}^{(3)}(k)$
correspond to time intervals for scaling limits of backbone paths.

The concept of the backbone is relevant for an understanding of
the value 6 of the upper critical dimension.  In the original lace expansion
for percolation \cite{HS90a}, as in the analysis involving the triangle
condition in \cite{AN84},
the leading behaviour corresponds to neglecting intersections between
a backbone and a percolation cluster.  Considering the backbone to
correspond to a 2-dimensional Brownian path, and the cluster to correspond
to a 4-dimensional ISE cluster, intersections will generically not occur
above $2+4=6$ dimensions.  This points to $d=6$ as the upper critical
dimension.

\section*{Acknowledgments}
This work was supported in part by NSERC, and was carried  
out in part while both authors were visiting
the University of British Columbia in 1997 and Microsoft Research in 1998.
The work of G.S.\ was also supported in part by a 1996
Invitation Fellowship of the Japan Society for the Promotion of Science.
We thank Michael Aizenman and Ed Perkins
for stimulating conversations and correspondence.

\providecommand{\bysame}{\leavevmode\hbox to3em{\hrulefill}\thinspace}


\begin{thebibliography}{10}

\bibitem{Aize97}
M.~Aizenman, \emph{On the number of incipient spanning clusters}, Nucl. Phys. B
  [FS] \textbf{{ 485}} (1997), 551--582.

\bibitem{AB87}
M.~Aizenman and D.J. Barsky, \emph{Sharpness of the phase transition in
  percolation models}, Commun. Math. Phys. \textbf{{ 108}} (1987),
  489--526.

\bibitem{AN84}
M.~Aizenman and C.M. Newman, \emph{Tree graph inequalities and critical
  behavior in percolation models}, J. Stat. Phys. \textbf{{ 36}} (1984),
  107--143.

\bibitem{Aldo93a}
D.~Aldous, \emph{The continuum random tree {III}}, Ann. Probab. \textbf{{
  21}} (1993), 248--289.

\bibitem{Aldo93}
\bysame, \emph{Tree-based models for random distribution of mass}, J. Stat.
  Phys. \textbf{{ 73}} (1993), 625--641.

\bibitem{BA91}
D.J. Barsky and M.~Aizenman, \emph{Percolation critical exponents under the
  triangle condition}, Ann. Probab. \textbf{{ 19}} (1991), 1520--1536.

\bibitem{BCKS98a}
C. Borgs, J.T.~Chayes, H.~Kesten, and J.~Spencer, \emph{The birth of the
infinite cluster:  finite size scaling in percolation}. In preparation.

\bibitem{BCKS98b}
C. Borgs, J.T.~Chayes, H.~Kesten, and J.~Spencer, \emph{Uniform boundedness
of critical crossing probabilities implies hyperscaling}. In preparation.

\bibitem{CCD87}
J.T. Chayes, L.~Chayes, and R.~Durrett, \emph{Inhomogeneous percolation
  problems and incipient infinite clusters}, J. Phys. A: Math. Gen.
  \textbf{{ 20}} (1987), 1521--1530.

\bibitem{DP97}
D.~Dawson and E.~Perkins, \emph{Measure-valued processes and renormalization of
  branching particle systems}, Stochastic Partial Differential Equations: Six
  Perspectives (R.~Carmona and B.~Rozovskii, eds.), AMS Math.\ Surveys and
  Monographs, 1997.

\bibitem{DS97}
E.~Derbez and G.~Slade, \emph{Lattice trees and super-{Brownian} motion}, 
  Canad.\ Math.\ Bull. \textbf{{ 40}} (1997), 19--38.

\bibitem{DS97a}
\bysame, \emph{The scaling limit of lattice trees in high
  dimensions}, Commun.\ Math.\ Phys. \textbf{{193}} (1998), 69--104.

\bibitem{Grim89}
G.~Grimmett, \emph{Percolation}, Springer, Berlin, 1989.

\bibitem{Grim96}
\bysame, \emph{Percolation and {Disordered} {Systems}}, St.\ Flour lecture
  notes, 1996.

\bibitem{HS98b}
T.~Hara and G.~Slade, \emph{The scaling limit of the incipient infinite cluster
  in high-dimensional percolation. {I}. {Critical} exponents}, In preparation.

\bibitem{HS98c}
T.~Hara and G.~Slade, \emph{The scaling limit of the incipient infinite cluster
  in high-dimensional percolation. {II}. {Integrated} super-{Brownian}
  excursion}, In preparation.

\bibitem{HS90a}
\bysame, \emph{Mean-field critical behaviour for percolation in high
  dimensions}, Commun. Math. Phys. \textbf{{ 128}} (1990), 333--391.

\bibitem{HS92c}
\bysame, \emph{The number and size of branched polymers in high dimensions}, J.
  Stat. Phys. \textbf{{ 67}} (1992), 1009--1038.

\bibitem{HS94}
\bysame, \emph{Mean-field behaviour and the lace expansion}, Probability and
  Phase Transition (Dordrecht) (G.\ Grimmett, ed.), Kluwer, 1994.

\bibitem{Hugh96}
B.D. Hughes, \emph{Random walks and random environments}, vol. 2: Random
  Environments, Oxford University Press, Oxford, 1996.

\bibitem{Kest82}
H.~Kesten, \emph{Percolation theory for mathematicians}, Birkh\"{a}user,
  Boston, 1982.

\bibitem{Kest86}
\bysame, \emph{The incipient infinite cluster in two-dimensional percolation},
  Probab. Th. Rel. Fields \textbf{{ 73}} (1986), 369--394.

\bibitem{LeGa93}
J.-F. Le~Gall, \emph{The uniform random tree in a {Brownian} excursion},
  Probab. Th. Rel. Fields \textbf{{ 96}} (1993), 369--383.

\bibitem{MS93}
N.~Madras and G.~Slade, \emph{The self-avoiding walk}, Birkh{\"a}user, Boston,
  1993.

\end{thebibliography}
\end{document}